\documentclass[lettersize,journal]{IEEEtran}
\usepackage{amsmath,amsfonts}
\usepackage{algorithmic}
\usepackage{algorithm}
\usepackage{array}
\usepackage[caption=false,font=normalsize,labelfont=sf,textfont=sf]{subfig}
\usepackage{textcomp}
\usepackage{stfloats}
\usepackage{url}
\usepackage{verbatim}
\usepackage{graphicx}
\usepackage{cite}
\usepackage{booktabs}
\usepackage{multirow}
\usepackage{threeparttable}
\usepackage{tikz}
\usepackage{hyperref}
\usetikzlibrary{shapes.geometric, arrows.meta, positioning, fit, backgrounds}

\begin{document}

\title{Probabilistic Fusion and Calibration of Neural Speaker Diarization Models}

\author{Juan Ignacio Alvarez-Trejos, Sergio A. Balanya, Daniel Ramos, Alicia Lozano-Diez}

\markboth{IEEE Preprint, 2025}
{Shell \MakeLowercase{\textit{et al.}}: A Sample Article Using IEEEtran.cls for IEEE Journals}


\maketitle
\begin{abstract}

End-to-End Neural Diarization (EEND) systems produce frame-level probabilistic speaker activity estimates, yet since evaluation focuses primarily on Diarization Error Rate (DER), the reliability and calibration of these confidence scores have been largely neglected. When fusing multiple diarization systems, DOVER-Lap remains the only established approach, operating at the segment level with hard decisions. We propose working with continuous probability outputs, which enables more sophisticated fusion and calibration techniques that can leverage model uncertainty and complementary strengths across different architectures. This paper presents the first comprehensive framework for calibrating and fusing EEND models at the probability level. We investigate two output formulations—multilabel and powerset representations—and their impact on calibration and fusion effectiveness. Through extensive experiments on the CallHome two-speaker benchmark, we demonstrate that proper calibration provides substantial improvements even for individual models (up to 19\% relative DER reduction), in some cases mitigating the absence of domain adaptation. We reveal that joint calibration in powerset space consistently outperforms independent per-speaker calibration, that fusion substantially improves over individual models, and that the Fuse-then-Calibrate ordering generally outperforms both calibrating before fusion and uncalibrated fusion while requiring calibration of only a single combined model. Our best configuration outperforms DOVER-Lap in terms of DER while providing reliable confidence estimates essential for downstream applications. This work proposes best practices for probability-level fusion of EEND systems and demonstrates the advantages of leveraging soft outputs over hard decisions.
\end{abstract}


\section{Introduction}

Speaker diarization identifies and temporally localizes individual speakers in multi-speaker audio recordings. Traditional modular diarization systems rely on assigning speech segments to speakers through a pipeline of speaker embedding extraction and clustering algorithms \cite{6135543}. While effective, these systems produce hard segment-level assignments where the distance to cluster centroids could provide some measure of uncertainty although it is not usually considered. To handle speaker overlap, external modules are often incorporated into the pipeline. The emergence of End-to-End Neural Diarization (EEND) \cite{fujita2019end} fundamentally changed this paradigm by directly predicting frame-level speaker activity probabilities for each time frame in an audio stream, enabling integrated modeling of overlapping speech and providing explicit confidence estimates at the frame level.

While recent EEND models achieve remarkably low error rates in speaker activity detection, both the reliability of their confidence estimates and the fusion of multiple diarization systems remain largely unexplored. Despite the critical role that confidence estimates play in downstream applications and model fusion, research on calibration of neural diarization models is extremely limited. The only prior work~\cite{plaquet24_interspeech} assesses calibration to select poorly calibrated samples for retraining, but does not explore post-hoc calibration techniques or model fusion. This gap becomes particularly critical when combining predictions from multiple models, as each model may exhibit distinct confidence patterns and biases that can significantly degrade fusion quality without proper calibration. The rapid evolution of EEND architectures has produced a diverse ecosystem of models with varying strengths: some excel at capturing long-range temporal dependencies \cite{10848627, eend-vector-clustering}, others at handling overlapping speech \cite{9688044}, and still others at speaker discrimination in noisy conditions \cite{10022924}.

This diversity naturally raises the question of whether these complementary capabilities can be effectively combined. Existing fusion approaches like DOVER \cite{9004031} and DOVER-Lap \cite{DoverLap21} operate on hard decisions at the segment level, limiting their ability to leverage the probabilistic nature of EEND outputs. We show that naive probabilistic model averaging is not able to effectively combine EEND models, and more advanced fusion techniques are required. We argue that this incompatibility arises from model \textit{miscalibration}---that is, their confidence predictions cannot be reliably interpreted as actual class probabilities \cite{Dawid, Brocker09, degroot83}. Modern neural networks (NNs), the backbone of EEND models, have shown to produce overconfident predictions~\cite{guo2017calibration, Ovadia2019}: for instance, when a network assigns 0.8 confidence to its predictions, the actual accuracy may be substantially lower than 80\%. Given that EEND systems provide frame-level probabilistic outputs, they are uniquely suited for calibration and fusion techniques that require continuous probability estimates---an approach not applicable to methods like DOVER-Lap that operate on hard segment-level decisions. 

This paper addresses these fundamental limitations towards a framework for calibrating and fusing neural diarization models. While post-hoc calibration techniques have been successfully applied in speaker recognition tasks~\cite{vanleeuwen2007introduction, 4291590}, their application to speaker diarization presents unique challenges. Our key insight is that the multilabel nature of speaker diarization, where multiple speakers can be simultaneously active in a time-frame, requires rethinking traditional calibration and fusion approaches designed for single-label classification problems. Moreover, well-calibrated probabilities enable optimal decision-making under varying costs and priors through decision theory—an important avenue for future work.

We investigate two output formulations for EEND speaker diarization: the standard multilabel formulation that treats speakers independently \cite{fujita2019end}, and the \textit{powerset} formulation that explicitly models one class for each of all speaker combinations \cite{du2022speaker, plaquet2023powerset}. Hereafter we will refer to these as Mult (Multilabel) and Power (Powerset) formulations. These output representations lead to different fusion behaviors and calibration requirements that have not been previously explored in neural diarization.

Through extensive experiments on the CallHome two-speaker benchmark \cite{martin2000nist}, we make several key contributions that represent, to our knowledge, the first systematic study of calibration and score-level fusion for EEND systems. First, we demonstrate that proper calibration provides substantial improvements even for individual EEND models, in some cases mitigating the absence of domain adaptation. Unlike prior work that operates on hard decisions, we show that calibrating frame-level probabilistic outputs enables more effective model combination. Second, we explore both unsupervised fusion methods and supervised approaches for combining EEND outputs, analyzing how each interacts with calibration strategies and domain adaptation. Third, we reveal that the choice between multilabel and powerset representations fundamentally impacts both calibration quality and fusion effectiveness, with certain methods benefiting from explicit modeling of speaker combinations while others performing similarly across both formulations. Importantly, we find that improvements in calibration quality (measured by Binary Cross-Entropy \cite{ferrer2024evaluating, e20030208}) generally translate to corresponding improvements in diarization performance (measured by Diarization Error Rate - DER), demonstrating that better-calibrated confidence estimates support more accurate multi-speaker segmentation. Finally, we propose best practices for score-level fusion of EEND speaker diarization systems, demonstrating that fusing first and then calibrating (Fuse-then-Calibrate) generally outperforms calibrating individual models before fusion, and that joint calibration in powerset space consistently outperforms independent per-speaker calibration.

The remainder of this paper is organized as follows. Section~\ref{sec:background} provides necessary background on EEND models, probability formulations, calibration, and fusion. Section~\ref{sec:methodology} presents our methodology for calibration and fusion of neural diarization models. Section~\ref{sec:experimental_setup} describes the experimental setup. Section~\ref{sec:results} presents comprehensive results analyzing calibration strategies, fusion methods, and their interactions. Section~\ref{sec:conclusion} concludes with key findings and future work.

\textbf{Code Availability:} Our calibration and fusion framework is publicly available at \href{https://github.com/SergioAlvarezB/calibrated-fusion-diarization}{\nolinkurl{github.com/SergioAlvarezB/calibrated-fusion-diarization}}, enabling researchers to apply these techniques to their own neural diarization systems.

\section{Background and Problem Formulation}
\label{sec:background}

\subsection{End-to-End Neural Diarization}
The EEND paradigm reformulates speaker diarization as a multilabel classification task in which NNs assign time frames to zero, one, or multiple speakers simultaneously~\cite{fujita2019end}. Since its introduction, EEND has evolved substantially with developments including self-attention mechanisms~\cite{eend_selfattention}, attractor-based approaches~\cite{horiguchi20_interspeech}, computational efficiency improvements~\cite{landini2024diaper}, and various architectural innovations~\cite{chen2023attention}.

Research has also explored diverse input feature representations for EEND models. Different feature types—including traditional acoustic features, speaker embeddings, and paralinguistic descriptors—have been shown to provide complementary information for speaker discrimination~\cite{app15094842, 10889475, 9746964}. This diversity in feature representations creates natural opportunities for score-level model fusion, as EEND models trained with distinct input features may capture different aspects of speaker characteristics and conversational dynamics.

In this work, we utilize pre-trained EEND with Encoder-Decoder-based Attractors (EEND-EDA)~\cite{horiguchi20_interspeech} models with different input feature representations as the foundation for our calibration and fusion framework. Following our approach in~\cite{app15094842}, we leverage multiple EEND-EDA models trained with diverse feature sets to exploit their complementary speaker information.

\subsection{Output Formulations in Neural Diarization}
While EEND established the multilabel classification paradigm as the foundation for neural diarization, recent research has explored alternative formulations that may better capture speaker interaction patterns by performing multi-class classification over the powerset of speakers~\cite{plaquet2023powerset}. The choice of a classification formulation has significant implications for both fusion strategies and calibration techniques.

\subsubsection{Multilabel Formulation}
The traditional EEND approach treats each speaker independently at the output level, producing speaker activity probabilities:
\begin{equation}
p_s^{\text{Mult}} \in [0,1], \quad s \in \{1, \ldots, S\}
\end{equation}
where $p_s^{\text{Mult}}$ represents the probability of speaker $s$ being active. Each $p_s^{\text{Mult}}$ is a probability from a one-vs-all binary classification problem, where $1 - p_s^{\text{Mult}}$ represents the probability that speaker $s$ is not active. These probabilities can be collected in a vector $\mathbf{p}^{\text{Mult}} = (p_1^{\text{Mult}}, \ldots, p_S^{\text{Mult}}) \in [0,1]^S$ for convenience.

While neural architectures such as self-attention mechanisms can capture dependencies between speakers within the model's internal representations, the final output layer treats each speaker prediction independently. Any speaker interaction patterns learned by the network remain implicit in the hidden representations rather than being explicitly modeled in the output formulation. This approach naturally handles overlapping speech but may not fully exploit speaker interaction information.

\subsubsection{Powerset Formulation}
An alternative approach considers all possible combinations of active speakers as mutually exclusive classes. For $S$ speakers, this creates $K = 2^S$ possible classes:
\begin{equation}
\mathcal{C} = \{\emptyset, \{1\}, \{2\}, \ldots, \{S\}, \{1,2\}, \ldots, \{1,2,\ldots,S\}\}
\end{equation}
The neural network outputs a probability for each class $c_k \in \mathcal{C}$:
\begin{equation}
p_{c_k}^{\text{Power}} \in [0,1], \quad k \in \{1, \ldots, K\}
\end{equation}
where these probabilities form a valid distribution: $\sum_{k=1}^{K} p_{c_k}^{\text{Power}} = 1$. These probabilities can be collected in a vector $\mathbf{p}^{\text{Power}} = (p_{c_1}^{\text{Power}}, \ldots, p_{c_K}^{\text{Power}}) \in \mathbb{S}^K$, where $\mathbb{S}^K$ denotes the $(K-1)$-dimensional probability simplex, the geometric locus where a $K$-dimensional probability vector can lie, i.e.:
\begin{equation}
\mathbb{S}^K = \left\{\mathbf{x} \in \mathbb{R}^K : \sum_{i=1}^{K} x_i = 1,\, x_i \geq 0\right\}
\end{equation}


\subsection{Post-hoc Calibration of Probabilistic Classifiers}

We use the term \textit{probabilistic classifier} to denote a classification system that outputs posterior probabilities or confidence scores. For a given input $\mathbf{x}$, the system outputs a prediction $\mathbf{\hat{p}}$ for the posterior class probabilities, where the $i$-th component represents the probability, or confidence, assigned to class $C_i$ such that $\hat{p}_i \simeq p(C_i|\mathbf{x})$. Let $\mathbf{y}$ denote the one-hot encoded target variable, where $y_i \in \{0, 1\}$ indicates whether $\mathbf{x}$ belongs to class $C_i$.

A probabilistic classifier is said to be well calibrated when its predicted probabilities accurately reflect the likelihood of correct predictions in terms of long-run observed frequencies. Formally, a classifier $f$ that produces confidence scores $f(\mathbf{x}) = \mathbf{\hat{p}}$ over $K$ possible classes, where $\mathbf{\hat{p}} \in \mathbb{S}^K$, is perfectly calibrated if:
\begin{equation}
P(\mathbf{y} | \mathbf{\hat{p}} = \mathbf{p}) = \mathbf{p}, \quad \forall \mathbf{p} \in \mathbb{S}^K
\end{equation}

\begin{figure}
    \centering
    \includegraphics[width=0.95\linewidth]{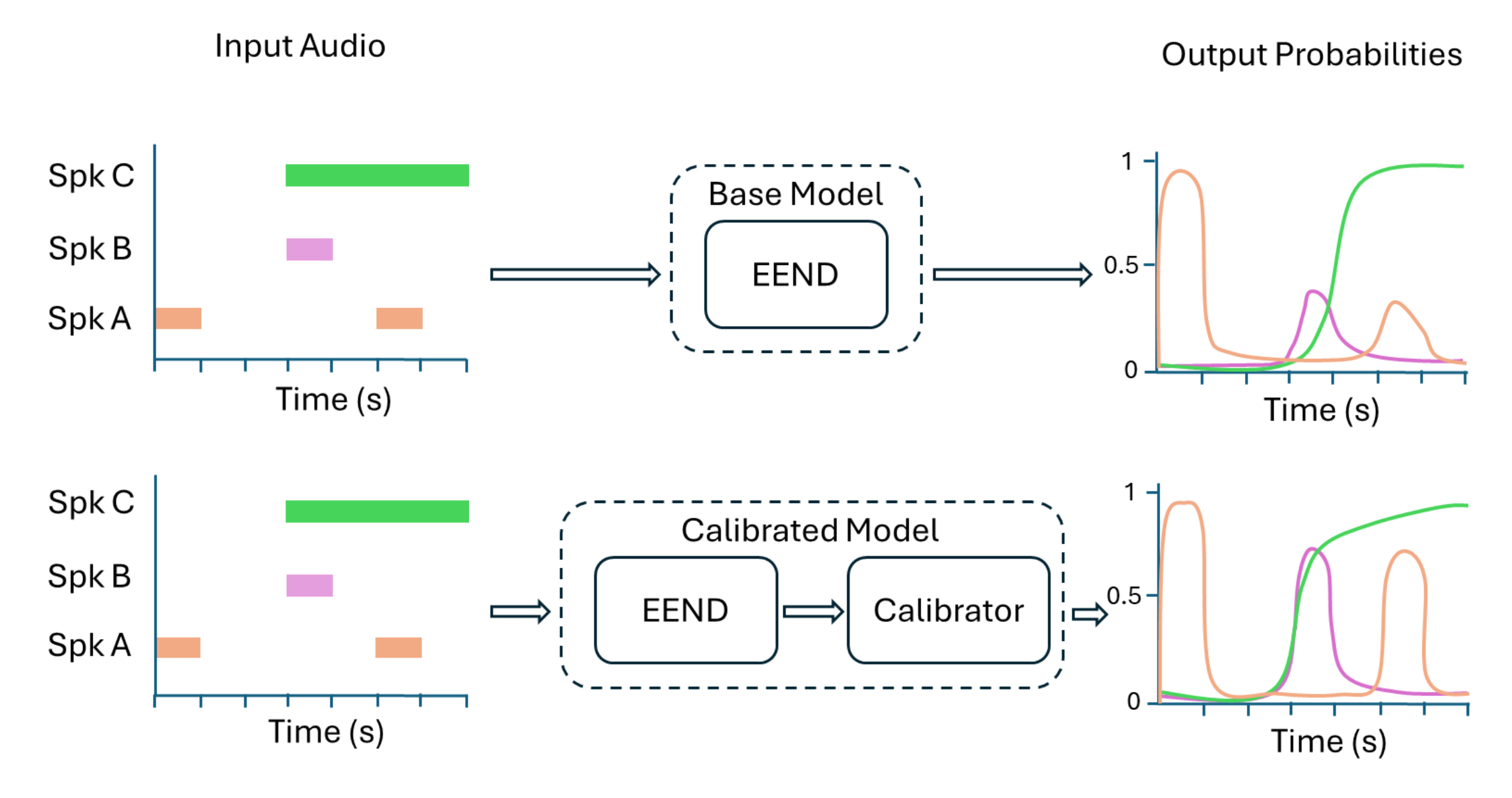}
    \caption{Post-hoc Calibration Process}
    \label{fig:post_hoc}
\end{figure}

Modern neural networks are typically not well calibrated out of the box~\cite{guo2017calibration, Ovadia2019}. A commonly used approach to address this issue is \textit{post-hoc calibration}, where the outputs of a miscalibrated classifier are recalibrated. This process involves training a secondary model (the calibration method) on the outputs of the primary classifier, as illustrated in Figure~\ref{fig:post_hoc}.

In this work, we employ Platt scaling~\cite{platt1999probabilistic} and its multi-class extension~\cite{niko_LRs}, which essentially applies flavours of logistic regression to the classifier outputs. The specific implementation of this regression varies depending on the calibration strategy, with detailed methods presented in the next section.

The general approach of Platt scaling, also known as matrix scaling, minimizes the cross-entropy loss over a calibration set with respect to learnable parameters. Let $\mathbf{z} \in \mathbb{R}^K$ denote the pre-softmax output vector, or logit vector, of a classifier over $K$ classes. The method is implemented as:
\begin{equation}
    \mathbf{\tilde{p}} = \text{softmax}\left(\mathbf{W}\mathbf{z} + \mathbf{b}\right)
\end{equation}
where $\mathbf{W} \in \mathbb{R}^{K\times K}$ and $\mathbf{b}\in \mathbb{R}^K$ are the learnable parameters, and $\mathbf{\tilde{p}}$ is the calibrated counterpart of $\mathbf{\hat{p}} = \text{softmax}(\mathbf{z}) = f(\mathbf{x})$.




Despite the importance of reliable confidence estimates in speaker diarization systems, research on calibration of neural diarization models remains extremely limited. To date, only one study has specifically addressed this topic, examining the calibration properties of powerset-based diarization models~\cite{plaquet24_interspeech}. This lack of research represents a significant gap, particularly considering the multilabel nature of the diarization task and the critical role that confidence estimates play in both downstream applications and model fusion scenarios.
\subsection{Model Fusion in Neural Diarization}

The diversity of neural architectures and feature representations in modern diarization systems creates opportunities for performance improvements through model fusion. Different models may excel in complementary scenarios: acoustic feature-based models might better capture phonetic distinctions, while speaker embedding models could provide superior speaker discrimination capabilities, and models with different temporal receptive fields may capture distinct speaker patterns.

Consider $M$ diarization models $f_1, f_2, \ldots, f_M$ producing predictions $\mathbf{p}_1, \mathbf{p}_2, \ldots, \mathbf{p}_M$ for the same input, where each $\mathbf{p}_m$ can be either a multilabel vector $\mathbf{p}_m \in [0,1]^S$ or a powerset distribution $\mathbf{p}_m \in \mathbb{S}^K$. The ensemble prediction can be formulated as:
\begin{equation}
\mathbf{p}_{\text{fused}} = g(\mathbf{p}_1, \mathbf{p}_2, \ldots, \mathbf{p}_M; \boldsymbol{\theta})
\end{equation}
where $g$ is the fusion function parameterized by $\boldsymbol{\theta}$.

Despite the potential benefits, model fusion in neural diarization remains largely underexplored compared to other speech processing tasks. The primary contribution to date is DOVER-Lap~\cite{DoverLap21}, which extends the original DOVER framework \cite{9004031} to handle overlapping segments through modified label mapping and voting mechanisms. DOVER-Lap operates by first aligning speaker labels across different system outputs using a global cost tensor that considers all pairwise overlaps simultaneously, followed by overlap-aware weighted majority voting that can assign multiple speakers to temporal regions.

However, DOVER-Lap still operates on hard decisions rather than probabilistic outputs, discarding valuable uncertainty information that could improve fusion quality.

\section{Methodology}
\label{sec:methodology}

This section presents our complete framework for fusing and calibrating multilabel neural diarization models. While we focus on EEND-EDA architectures \cite{horiguchi20_interspeech}, our methodology is applicable to any end-to-end neural diarization system that produces frame-level speaker activity probabilities in multilabel format.

\subsection{Probability Space Transformation}
\label{subsec:space_transformation}

Given the multilabel and powerset formulations presented above, we can transform between these probability spaces to enable flexible fusion and calibration strategies. Starting from multilabel predictions $\mathbf{p}^{Mult} = [p_{s_1}, p_{s_2}]$, where $p_{s_i}$ represents the probability of speaker $s_i$ being active, the transformation to powerset space assumes independence between speakers: $P(s_1, s_2) = P(s_1)P(s_2)$. Under this assumption, the probability of each possible speaker activity combination is given by:

\begin{equation}
\mathbf{p}^{Power} = \begin{bmatrix}
(1-p_{s_1})(1-p_{s_2}) \\
p_{s_1}(1-p_{s_2}) \\
(1-p_{s_1})p_{s_2} \\
p_{s_1} p_{s_2}
\end{bmatrix}
\end{equation}

While the independence assumption ignores any speaker dependencies learned by the neural network---as these are implicit in the multilabel predictions and cannot be recovered during the transformation---calibration methods applied in the powerset space may partially compensate for this limitation by learning to model speaker interactions through the calibration parameters.

\subsection{Model Fusion Strategies}
\label{subsec:fusion_strategies}

We investigate multiple fusion approaches that combine predictions from several diarization models, ranging from simple averaging to sophisticated confidence-weighted schemes. Let $\mathbf{p}_m$ denote the probability predictions from model $m \in \{1,2,...,M\}$, where $M=3$ in our experiments, and $\mathbf{z}_m$ denote the corresponding logits. These predictions can be represented in either multilabel space ($\mathbf{p}_m \in [0,1]^S$, $\mathbf{z}_m \in \mathbb{R}^S$) or powerset space ($\mathbf{p}_m \in [0,1]^K$, $\mathbf{z}_m \in \mathbb{R}^K$), depending on the output formulation. These fusion strategies can be broadly categorized into two groups: unsupervised methods that directly combine model outputs without requiring additional training data, and supervised methods that learn combination weights from labeled examples.

\subsubsection{Unsupervised Fusion Methods}

We explore four unsupervised fusion strategies to combine model outputs based on different principles:

\begin{itemize}
\item \textbf{Average Probabilities} represents the most straightforward fusion approach, computing the arithmetic mean of model predictions:
\begin{equation}
\mathbf{p}_{fused} = \frac{1}{M} \sum_{m=1}^{M} \mathbf{p}_m
\end{equation}

This method assumes that all models contribute equally to the decision and provides a stable baseline for comparison.

\item \textbf{Average Logits} performs fusion in the logit space before applying the appropriate activation function (also known as the inverse-link function~\cite{mccullagh1984generalized}):
\begin{equation}
\mathbf{p}_{\text{fused}} = \phi\left(\frac{1}{M} \sum_{m=1}^{M} \mathbf{z}_m\right)
\end{equation}
where $\phi$ is the activation function: sigmoid $\phi(\mathbf{z})_s = \sigma(z_s) = \frac{1}{1 + \exp(-z_s)}$ for multilabel, or softmax $\phi(\mathbf{z})_k = \frac{\exp(z_k)}{\sum_{k'=1}^{K} \exp(z_{k'})}$ for powerset. This approach handles extreme probability values more effectively and preserves the relative confidence differences between models.

\item \textbf{Dynamic Logits Fusion} introduces adaptive weighting based on prediction confidence:
\begin{equation}
\mathbf{p}_{\text{fused}} = \phi\left(\sum_{m=1}^{M} w_m \cdot \mathbf{z}_m\right)
\end{equation}
where weights $w_m$ are computed based on the absolute sum of logits as a confidence measure:
\begin{equation}
w_m = \frac{\sum_{i} |z_{m,i}|}{\sum_{k=1}^{M} \sum_{i} |z_{k,i}|}
\end{equation}
where $i$ indexes speakers (multilabel) or classes (powerset). This approach addresses the fact that different models may produce logits at different scales. By normalizing based on logit magnitudes, models with higher confidence (larger absolute logits) receive greater weight in the fusion, allowing the system to dynamically prioritize more certain predictions while accounting for scale differences across models.

\item \textbf{Entropy Fusion} employs inverse entropy weighting, where models with lower predictive entropy contribute more to the final decision:
\begin{equation}
\mathbf{p}_{\text{fused}} = \sum_{m=1}^{M} w_m \cdot \mathbf{p}_m
\end{equation}
where the weights are computed as:
\begin{equation}
w_m = \frac{H_{\max} - H(\mathbf{p}_m)}{\sum_{k=1}^{M} (H_{\max} - H(\mathbf{p}_k))}
\end{equation}
and $H(\mathbf{p}_m) = -\sum_i p_{m,i} \log p_{m,i}$ is the predictive entropy, with $i$ indexing speakers (multilabel) or classes (powerset), and $H_{\max} = \log S$ (multilabel) or $H_{\max} = \log K$ (powerset) denoting the maximum possible entropy for uniform distributions.

\end{itemize}

\subsubsection{Supervised Fusion Method}
In contrast to the unsupervised approaches, we investigate a data-driven fusion strategy using a so-called \textbf{MetaLearner}—a logistic regression model that learns to combine system outputs. The MetaLearner learns optimal combination weights from labeled data, taking the concatenated logits from all models as input:
\begin{equation}
    \mathbf{z}_{\text{fused}} = \mathbf{W}[\mathbf{z}_1; \mathbf{z}_2; \ldots; \mathbf{z}_M] + \mathbf{b}
\end{equation}
where $[\cdot;\cdot]$ denotes concatenation, and $\mathbf{W} \in \mathbb{R}^{D \times (M \cdot D)}$ and $\mathbf{b} \in \mathbb{R}^D$ are learned parameters optimized to minimize cross-entropy loss on the training set, with $D = S$ for multilabel or $D = K$ for powerset. This supervised approach can capture complex relationships between model outputs that may not be apparent in unsupervised methods.




\subsection{Calibration Framework}
\label{subsec:calibration_framework}

Neural models often produce poorly calibrated predictions~\cite{guo2017calibration}, where the predicted probabilities do not accurately reflect the empirical proportions of events~\cite{Brocker09, Dawid}. This miscalibration can significantly impact fusion performance, as methods that rely on probability magnitudes may be misled by systematic biases in individual models.

We implement post-hoc calibration using logistic regression, also known as Platt Scaling \cite{platt1999probabilistic}, which has proven effective in other speech multiclass tasks~\cite{niko_LRs}. The rationale behind logistic regression calibration relies on its objective function, the cross-entropy, which is a proper scoring rule and its minimization improves calibration~\cite{niko_LRs}. For the case when we have independent scores per speaker, i.e. the multilabel domain, we explore two calibration strategies: Independent and Joint Calibration. For the powerset domain, only joint calibration is applicable, as the output represents a probability distribution over mutually exclusive classes.

\textbf{Independent Calibration} treats each class separately, learning individual transformation parameters for each speaker:
\begin{equation}
p^{cal}_{i} = \sigma(\alpha_i \log(p_i) + \beta_i)
\end{equation}
where $\alpha_i$ and $\beta_i$ are class-specific parameters learned independently for each class $i$. This approach allows for class-specific correction of calibration biases but ignores potential dependencies between classes. 


\textbf{Joint Calibration} learns a unified transformation that calibrates all classes simultaneously. For multilabel scenarios, the sigmoid is applied element-wise:
\begin{equation}
\mathbf{p}^{cal} = \sigma(\mathbf{A} \log(\mathbf{p}) + \mathbf{b})
\end{equation}
where $\mathbf{A} \in \mathbb{R}^{K\times K}$ is the learned transformation matrix and $\mathbf{b} \in \mathbb{R}^K$ is a bias vector. For multiclass (powerset) scenarios:
\begin{equation}
\mathbf{p}^{cal} = \text{softmax}(\mathbf{W} \log(\mathbf{p}) + \mathbf{b})
\end{equation}
where the softmax ensures valid probability distributions, i.e. $\mathbf{p}^{cal} \in \mathbb{S}^K$. 

\subsection{Modular Framework Design}
\label{subsec:integration_strategy}


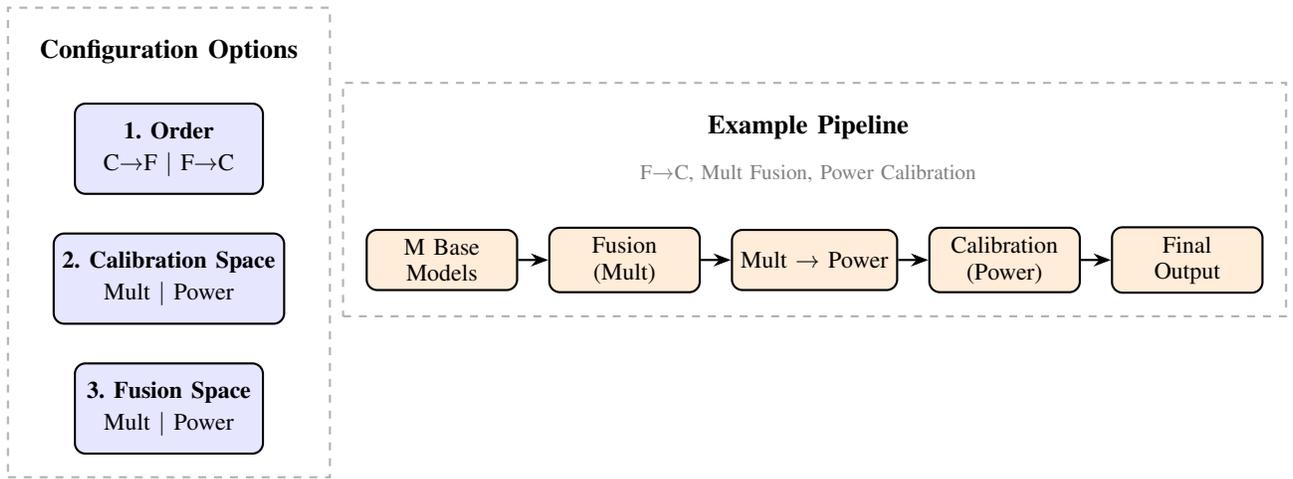
\begin{figure*}[htbp]
\centering
\begin{tikzpicture}[
    node distance=0.6cm,
    decision/.style={rectangle, rounded corners, minimum width=2.5cm, minimum height=1.2cm, 
                     text centered, draw=black, fill=blue!10, thick, align=center, font=\small},
    process/.style={rectangle, rounded corners, minimum width=2cm, minimum height=0.8cm,
                    text centered, draw=black, fill=orange!15, thick, font=\small},
    arrow/.style={-Stealth, thick},
    label/.style={font=\bfseries},
    dashed-box/.style={dashed, thick, draw=gray!60}
]

\node[label] (config-title) at (0, 0) {Configuration Options};

\node[decision, below=0.4cm of config-title] (decision1) {
    \textbf{1. Order}\\[2pt]
    C→F \textbar\ F→C
};

\node[decision, below=0.5cm of decision1] (decision2) {
    \textbf{2. Calibration Space}\\[2pt]
    Mult \textbar\ Power
};

\node[decision, below=0.5cm of decision2] (decision3) {
    \textbf{3. Fusion Space}\\[2pt]
    Mult \textbar\ Power
};

\begin{scope}[on background layer]
    \node[fit=(config-title)(decision1)(decision2)(decision3), 
          dashed-box, inner sep=0.3cm] (config-box) {};
\end{scope}

\node[label] (pipeline-title) at (8.5, -1) {Example Pipeline};
\node[below=0.1cm of pipeline-title, font=\footnotesize, text=gray] (example-note) 
    {F→C, Mult Fusion, Power Calibration};

\node[process, below left=0.5cm and 1.5 of example-note,align=center] (base) {M Base\\Models};
\node[process, right=0.4cm of base, align=center] (fusion) {Fusion\\(Mult)};
\node[process, right=0.4cm of fusion, align=center] (transform) {Mult → Power};
\node[process, right=0.4cm of transform, align=center] (calibration) {Calibration\\(Power)};
\node[process, right=0.4cm of calibration, align=center] (output) {Final\\Output};

\draw[arrow] (base) -- (fusion);
\draw[arrow] (fusion) -- (transform);
\draw[arrow] (transform) -- (calibration);
\draw[arrow] (calibration) -- (output);

\begin{scope}[on background layer]
    \node[fit=(pipeline-title)(example-note)(base)(fusion)(transform)(calibration)(output), 
          dashed-box, inner sep=0.3cm] (pipeline-box) {};
\end{scope}

\end{tikzpicture}
\caption{Modular framework design with three independent configuration decisions (left) and an example horizontal processing pipeline (right). The framework allows independent selection of: (1) calibration-fusion ordering (C→F: Calibrate-then-Fuse or F→C: Fuse-then-Calibrate), (2) calibration probability space (Mult: Multilabel or Power: Powerset), and (3) fusion probability space. The example pipeline demonstrates the F→C strategy with fusion in multilabel space followed by calibration in powerset space.}
\label{fig:modular_diagram}
\end{figure*}

Our framework provides a modular approach to combining calibration and fusion for neural diarization models. The design encompasses three key decisions that can be independently configured:

\begin{enumerate}
\item \textbf{Calibration-Fusion Order}: Whether to first calibrate individual base models and then fuse their predictions, or to first fuse the base model predictions and then calibrate the combined output.

\item \textbf{Calibration Space}: Whether to perform calibration in the multilabel or powerset probability space.

\item \textbf{Fusion Space}: Whether to perform fusion in the multilabel or powerset probability space.
\end{enumerate}

This modular design allows for flexible exploration of different configurations. Probability space transformations enable calibration and fusion to be performed in different spaces, allowing us to identify the optimal combination of ordering, representation, and methodology for each component.

The framework also enables investigation of whether supervised fusion methods like the MetaLearner inherently perform calibration during training, potentially reducing the need for explicit calibration steps.

\section{Experimental Setup}
\label{sec:experimental_setup}

\subsection{Base Diarization Models and Conditions}
\label{subsec:base_models}

We evaluate our framework using three EEND-EDA models previously developed for two-speaker diarization \cite{app15094842}. All models share the same EEND-EDA architecture: a 4-block encoder with 256-dimensional outputs, trained on 50-second segments with a maximum of 15 attractors (though only 2 are actively trained for two-speaker scenarios). Complete implementation details are provided in the cited work; here we highlight the key differences in their input features:
\begin{itemize}
\item \textbf{MFB} utilizes 23-dimensional Mel-filterbank features with frame stacking. Features are extracted using a 25ms window with 10ms shift, then contextualized by concatenating 7 previous and 7 subsequent frames, resulting in a 345-dimensional input vector (23 × 15 frames).
\item \textbf{ECAPA-TDNN $\oplus$ MFB} enhances the MFB model by concatenating 512-dimensional ECAPA-TDNN speaker embeddings extracted with 1-second windows and 100ms shift with the 345-dimensional MFB features, creating a 857-dimensional input vector. This configuration leverages specialized speaker-discriminative representations alongside traditional acoustic features.
\item \textbf{GeMAPS $\oplus$ MFB} incorporates paralinguistic features by combining a reduced 52-parameter GeMAPS feature set with Mel-filterbank features. The GeMAPS features are extracted using 60ms windows with 10ms shift and contextualized with 2 frames on each side (5 total frames), resulting in a 260-dimensional vector (52 × 5 frames) that is concatenated with the 345-dimensional MFB features for a final 605-dimensional input.
\end{itemize}
Each model is evaluated under two training conditions to assess the impact of domain adaptation:

\begin{itemize}
\item \textbf{No Fine-tuning (No FT)} represents models trained exclusively on simulated conversations, without adaptation to the target domain. This condition allows us to investigate whether calibration can compensate for domain mismatch.

\item \textbf{Fine-tuned (FT)} includes models that underwent domain adaptation to CallHome using the CH1 two-speaker subset with Adam optimizer (learning rate 1e-4) following the original EEND-EDA methodology \cite{horiguchi20_interspeech}.
\end{itemize}

\subsection{Dataset and Evaluation Protocol}
\label{subsec:dataset}

\textbf{Training Data:} All base models are initially trained on simulated conversations (SC) generated following established methodologies \cite{landini22_interspeech}. The dataset comprises 2,480 hours of two-speaker conversations created using recordings from multiple corpora: Switchboard-2 (Phases I, II, and III), Switchboard Cellular (Parts 1 and 2) \cite{godfrey1992switchboard}, and NIST Speaker Recognition Evaluation datasets (2004, 2005, 2006, and 2008) \cite{912681, 4013537, martin2009nist}. All source material is standardized to 8 kHz sampling rate to match telephone speech conditions. The simulated conversations include background noise from the MUSAN dataset \cite{snyder2015musan} and room impulse responses from the RIR dataset \cite{ko2017study} to enhance robustness.

\textbf{Evaluation Data:} All experiments are conducted on the CallHome corpus \cite{martin2000nist}, a widely-used benchmark for conversational telephone speech diarization. We focus on the two-speaker subset of the CH2 test set, which contains 148 recordings with approximately 3 hours of total audio.

\textbf{Training and Calibration Protocol:} Calibration parameters and MetaLearner weights are estimated using model predictions on the CallHome two-speaker subset of the CH1 adaptation set, which serves as the training data for both calibration methods and supervised fusion. Note that this is the same dataset in which the base models are fine-tuned.


\subsection{Implementation Details}
\label{subsec:implementation}

\textbf{Model Processing:} All models process input sequences with a subsampling factor of 10, reducing the temporal resolution from 10ms to 100ms frames for computational efficiency. During inference, predictions are upsampled back to the original temporal resolution for accurate boundary detection.

\textbf{Post-processing:} Model outputs undergo median filtering with an 11-frame window (corresponding to 110ms at the upsampled resolution) to smooth predictions and reduce spurious activations. Final speaker decisions are made using a fixed threshold of 0.5 applied to the filtered probabilities. While calibration enables optimal decision-making in a richer decision-theoretic framework with application-specific costs and utilities, exploring such frameworks is beyond the scope of this work. Here, calibration primarily serves to improve fusion quality by ensuring that probability estimates from different models are comparable and can be meaningfully combined.

\textbf{Platt Scaling:} The training procedure is identical across all implementations of Platt Scaling, whether used for calibration or fusion. Parameters are optimized by minimizing cross-entropy loss using the L-BFGS (Limited-memory Broyden-Fletcher-Goldfarb-Shanno) algorithm with default L2 regularization (C=1.0) and a maximum of 1000 iterations.


\textbf{Fusion Processing:} All fusion experiments are conducted at the 100ms frame level before applying post-processing steps. This ensures that fusion decisions are made on the same temporal granularity across all methods.

\subsection{Evaluation Metrics}
\label{subsec:metrics}

We employ two complementary metrics that assess both diarization performance and prediction quality:

\textbf{Diarization Error Rate (DER)} serves as our primary task-specific metric and the standard evaluation measure in speaker diarization. It measures the percentage of time frames with incorrect speaker attribution, including missed speech, false alarms, and speaker confusion errors.

\textbf{Cross-Entropy (CE)} is a proper scoring rule that measures the overall quality of probabilistic predictions, considering both calibration and discriminative power. For multilabel classification, CE is computed as Binary Cross-Entropy (BCE) per speaker. For powerset representations, we transform predictions back to multilabel space and compute BCE to maintain a consistent evaluation metric across both formulations. Well-calibrated models do not necessarily achieve lower CE scores, as a model can be perfectly calibrated while having poor discrimination (e.g., by predicting the prior distribution). Therefore, it is important to measure overall prediction quality with a proper scoring rule like cross-entropy~\cite{ferrer2024evaluating}. However, post-hoc calibration improves prediction quality by mainly reducing calibration error~\cite{Brocker09}.



\section{Framework Analysis and Results}
\label{sec:results}

This section presents a comprehensive analysis of our calibration and fusion framework across multiple dimensions. First, we examine the impact of calibration on the performance of individual base models and their fusion. Second, we compare joint calibration of all speakers versus independent per-speaker calibration. Third, we investigate how the choice of probability space affects both calibration and fusion performance. Fourth, we analyze the effects of different calibration-fusion ordering strategies. Finally, we provide a visual analysis of the impact that fusion and calibration have on each DER component and on the BCE. 

Before presenting our results, it is important to note that the base models employed in this work do not represent the current state-of-the-art in speaker diarization. More advanced architectures such as SortFormer-Hybrid-Loss with 123M parameters achieve 5.87\% DER on CallHome 2-speaker~\cite{park2025sortformer}, and recent EEND variants like AED-EEND-EE with 11.6M parameters reach 6.93\% DER~\cite{10438838}. However, the goal of this work is not to achieve state-of-the-art diarization performance, but rather to systematically investigate calibration and fusion strategies that can be applied to any neural diarization system. The performance of our base models provides a clearer view of how calibration and fusion techniques contribute to overall system performance.

\subsection{Calibration of Base Models}
In this subsection, we present a baseline case that demonstrates the effect of calibrating base models before combining their predictions. First, base models are calibrated with Platt Scaling in the powerset domain following the procedure detailed in Section \ref{subsec:calibration_framework}. Then their predictions are combined in the multilabel space using the different fusion strategies described in the previous section. Finally, performance of each combination is measured in terms of DER and BCE.

Table \ref{tab:exp_1_combined} shows performance metrics for models without and with fine-tuning, before and after calibration. The fused models showcased here were obtained using the simple Average Probs strategy, a common approach for combining predictions.

\begin{table}[htbp]
\centering
\caption{DER (\%) and BCE performance of base models and their fusion before and after calibration. Results shown for models without (No FT) and with fine-tuning (FT).}
\label{tab:exp_1_combined}
\begin{tabular}{llcccc}
\toprule
\textbf{Training} & \textbf{Model} & \multicolumn{2}{c}{\textbf{DER (\%)}} & \multicolumn{2}{c}{\textbf{BCE}} \\
\cmidrule(lr){3-4} \cmidrule(lr){5-6}
\textbf{Condition} &  & \textbf{Before} & \textbf{After} & \textbf{Before} & \textbf{After} \\
\midrule
\multirow{4}{*}{\textbf{No FT}} 
& MFB            & 10.381 & 8.397 & 0.419 & 0.271 \\
& ECAPA-TDNN     & 11.136 & 10.370 & 0.551 & 0.322 \\
& GeMAPS         & 10.780 & 9.467 & 0.416 & 0.297 \\
\cmidrule{2-6}
& Fused (Avg Probs) & 8.425 & 7.764 & 0.257 & 0.252 \\
\midrule
\multirow{4}{*}{\textbf{FT}} 
& MFB            & 8.236 & 7.834 & 0.310 & 0.258 \\
& ECAPA-TDNN     & 9.054 & 9.036 & 0.436 & 0.290 \\
& GeMAPS         & 9.068 & 8.988 & 0.288 & 0.263 \\
\cmidrule{2-6}
& Fused (Avg Probs) & 7.067 & 7.031 & 0.214 & 0.213 \\
\bottomrule
\end{tabular}
\end{table}

The results show that individual base models benefit from calibration, especially those without fine-tuning. Since the calibration set coincides with the fine-tuning set, calibration provides not only better-calibrated probabilities but also a degree of domain adaptation. For example, MFB improves from 10.381\% to 8.397\% DER (19.1\% relative reduction), while BCE improves from 0.419 to 0.271. After calibration, non-fine-tuned models achieve performance comparable to fine-tuned models: calibrated MFB (8.397\% DER) approaches fine-tuned uncalibrated MFB (8.236\% DER). This effect becomes more pronounced when fusing predictions: the fusion of calibrated models provides a 7.8\% relative improvement in DER over raw fusion for non-fine-tuned models (8.425\% to 7.764\%), but only 0.5\% for fine-tuned ones (7.067\% to 7.031\%). Nonetheless, calibration consistently improves performance across all model configurations, with BCE improvements ranging from 0.025 (FT GeMAPS) to 0.229 (No FT ECAPA-TDNN).

\begin{figure}[h]
  \centering
  \includegraphics[width=\linewidth]{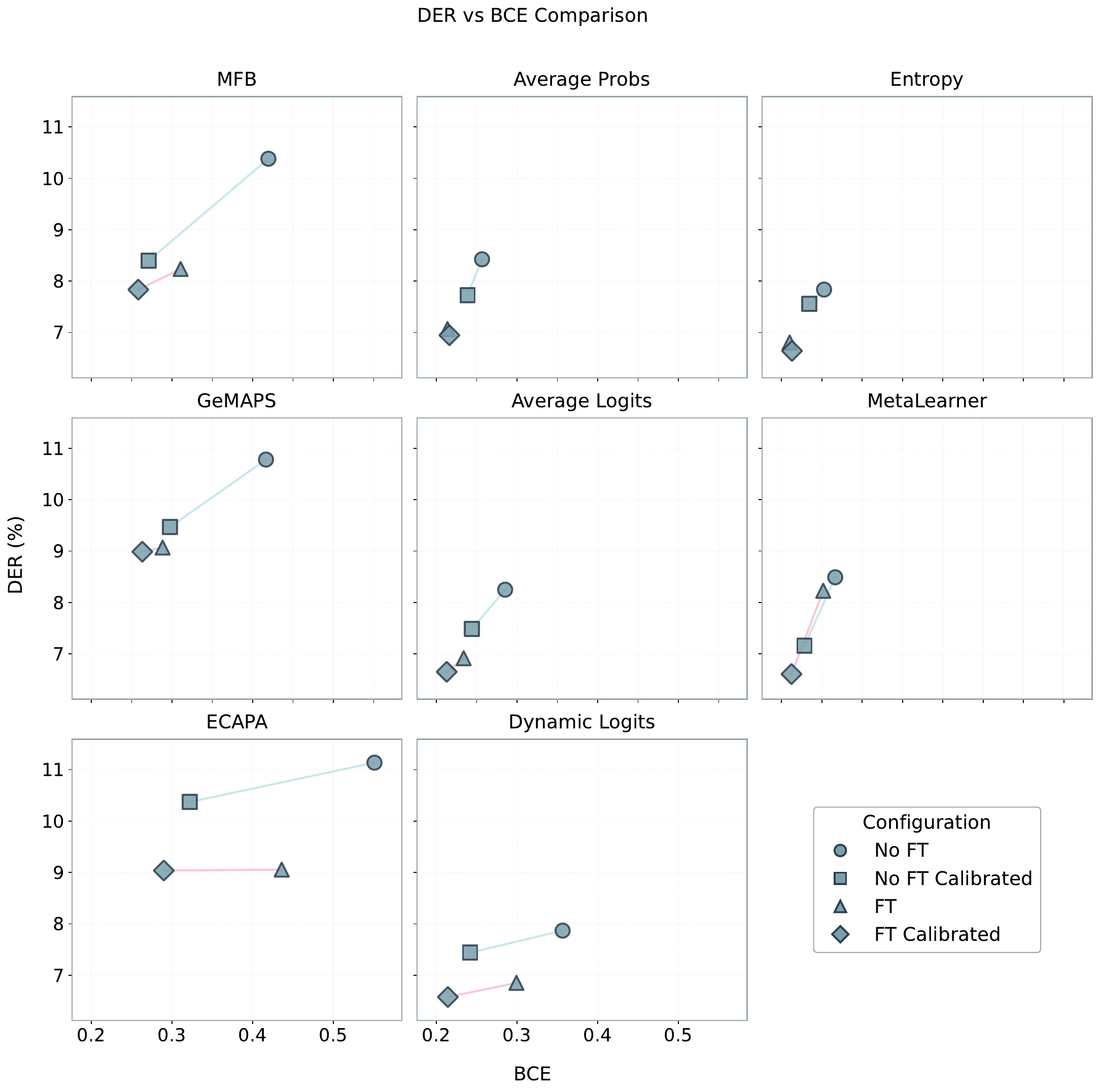}
  \caption{DER (\%) vs BCE comparison on Powerset space for different models and fusion methods.}
  \label{fig:derbce}
\end{figure}

Figure~\ref{fig:derbce} presents a more comprehensive comparison including all fusion strategies. The observation that non-fine-tuned models benefit more from calibration holds across all fusion methods. The figure also demonstrates that improvements in BCE resulting from calibration are almost always accompanied by corresponding improvements in DER. The only notable exception is the fine-tuned ECAPA model, which exhibits peculiar behavior likely attributable to the use of oracle VAD during training~\cite{app15094842}. By setting feature vectors to zero during non-speech segments, oracle VAD alters the distribution of the model's output scores, potentially affecting the relationship between calibration and task performance. However, a detailed analysis of this phenomenon is beyond the scope of this work.

\subsection{Joint versus Independent Calibration Strategies}
This section compares the two calibration strategies applicable to the multilabel probability space: joint calibration versus independent per-speaker calibration. Following the calibration-then-fusion approach, we first calibrate the non-fine-tuned base models in the multilabel domain, then fuse their predictions and compare performance across both calibration strategies.

\begin{table}[htbp]
\centering
\caption{DER (\%) and BCE performance of base models and their fusion after independent (Indep) and joint calibration in the multilabel space. Results shown for models without fine-tuning.}
\label{tab:exp_2}
\begin{tabular}{lcccc}
\toprule
\textbf{Model} & \multicolumn{2}{c}{\textbf{DER (\%)}} & \multicolumn{2}{c}{\textbf{BCE}} \\
\cmidrule(lr){2-3} \cmidrule(lr){4-5}
 & \textbf{Indep} & \textbf{Joint} & \textbf{Indep} & \textbf{Joint} \\
\midrule
MFB            & 12.059 & 10.874 & 0.351 & 0.329 \\
ECAPA-TDNN     & 17.724 & 12.363 & 0.514 & 0.360 \\
GeMAPS         & 14.037 & 11.883 & 0.378 & 0.354 \\
\midrule
Average Probs  & 9.674 & 8.251 & 0.315 & 0.284 \\
Average Logits & 8.672 & 7.698 & 0.309 & 0.266 \\
Dynamic Logits & 8.313 & 7.342 & 0.365 & 0.263 \\
Entropy        & 8.453 & 7.533 & 0.291 & 0.264 \\
MetaLearner    & 7.535 & 7.534 & 0.252 & 0.252 \\
\midrule
DOVER-Lap\textsuperscript{*} & 11.440 & 10.610 & - & - \\
\bottomrule
\end{tabular}
\vspace{2mm}

\footnotesize *DOVER-Lap operates at segment-level fusion using MFB as primary system.
\end{table}

Table~\ref{tab:exp_2} shows that joint calibration almost consistently outperforms independent per-speaker calibration across both base models and fusion methods. The improvements are substantial for individual models: MFB improves from 12.059\% to 10.874\% DER (9.8\% relative reduction), ECAPA-TDNN from 17.724\% to 12.363\% DER (30.2\% relative reduction), and GeMAPS from 14.037\% to 11.883\% DER (15.3\% relative reduction). Fusion methods also benefit considerably, with DER improvements ranging from 0.920\% (Entropy: 8.453\% to 7.533\%) to 1.423\% (Average Probs: 9.674\% to 8.251\%). BCE improvements follow similar patterns, with reductions of 0.022 to 0.154 for individual models and 0.027 to 0.102 for fusion methods. The MetaLearner shows minimal change, performing similarly under both calibration strategies (7.535\% vs 7.534\% DER, 0.252 BCE for both).

This finding is particularly significant because it reveals that speaker dependencies exist even in the multilabel formulation, where speakers are treated independently at the output level. While the multilabel representation assumes independence between speakers, the joint calibration strategy can exploit underlying dependencies in the prediction errors to improve calibration quality. This suggests that calibration should account for inter-speaker relationships, and that joint calibration strategies are preferable when operating in the multilabel space, as they can capture and correct for systematic biases that affect multiple speakers simultaneously. Based on this finding, all subsequent experiments in this work employ joint calibration.

\subsection{Comparison of Fusion Methods Across Probability Spaces}

Table~\ref{tab:fusion_comparison} presents a comparison of fusion methods across both probability spaces and training conditions, including comparison with the existing DOVER-Lap~\cite{DoverLap21} fusion approach.

\begin{table}[htbp]
\centering
\caption{Performance comparison of fusion methods in multilabel (Mult) versus powerset (Power) probability spaces. Results show DER (\%) and BCE for both non-fine-tuned (No FT) and fine-tuned (FT) models.}
\label{tab:fusion_comparison}
\resizebox{\columnwidth}{!}{%
\begin{tabular}{llcccc}
\toprule
\textbf{Training} & \textbf{Method} & \multicolumn{2}{c}{\textbf{DER (\%)}} & \multicolumn{2}{c}{\textbf{BCE}} \\
\cmidrule(r){3-4} \cmidrule(l){5-6}
\textbf{Condition} &  & \textbf{Mult} & \textbf{Power} & \textbf{Mult} & \textbf{Power} \\
\midrule
\multirow{6}{*}{\textbf{No FT}} 
& Average Probs  & 8.425 & 8.425 & 0.257 & 0.257 \\
& Average Logits & 8.248 & 8.248 & 0.286 & 0.285 \\
& Dynamic Logits & 7.811 & 7.867 & 0.375 & 0.356 \\
& Entropy        & 8.428 & 7.835 & 0.269 & 0.253 \\
& MetaLearner    & \textbf{7.326} & 8.490 & \textbf{0.234} & 0.267 \\
\cmidrule{2-6}
& DOVER-Lap & 9.030 & -- & -- & -- \\
\midrule
\multirow{6}{*}{\textbf{FT}} 
& Average Probs  & 7.067 & 7.067 & 0.214 & 0.214 \\
& Average Logits & 6.915 & 6.915 & 0.234 & 0.234 \\
& Dynamic Logits & 6.830 & 6.850 & 0.314 & 0.299 \\
& Entropy        & 6.910 & \textbf{6.806} & 0.220 & \textbf{0.210} \\
& MetaLearner    & 6.963 & 8.227 & 0.217 & 0.252 \\
\cmidrule{2-6}
& DOVER-Lap & 7.030 & -- & -- & -- \\
\bottomrule
\end{tabular}
}

\end{table}

All fusion methods provide substantial improvements in DER over the best individual model (8.236\% DER for fine-tuned MFB). The existing DOVER-Lap method achieves 7.030\% DER with fine-tuned models, which serves as a competitive baseline but is outperformed by most of our proposed probability-level fusion approaches.

Comparing performance across probability spaces, the results show no clear systematic advantage of fusing in the powerset domain versus the multilabel domain. While some methods (e.g., Entropy fusion) show slight improvements in powerset space, others (e.g., MetaLearner) perform substantially better in multilabel space. Given this lack of consistent advantage and the fact that the multilabel space has significantly lower dimensionality ($S$ versus $K=2^S$ classes), fusion in the multilabel domain is preferable for better computational efficiency and comparable performance.

\subsection{Impact of Calibration Probability Space}

This section examines the impact of the probability space on calibration performance. Base models are first calibrated in either the multilabel or powerset space and then fused in the multilabel space. To account for possible interactions with the fusion strategy, we report results for all fusion methods.

\begin{table}[htbp]
\centering
\caption{Impact of calibration space on non-fine-tuned model performance. DER (\%) and BCE are compared after calibration in multilabel (Mult) versus powerset (Power) spaces for both individual models and fusion methods.}
\label{tab:calibration_probability_spaces}
\begin{tabular}{lcccc}
\toprule
 & \multicolumn{2}{c}{\textbf{DER (\%)}} & \multicolumn{2}{c}{\textbf{BCE}}\\
\cmidrule(lr){2-3} \cmidrule(lr){4-5}
\textbf{Method} & \textbf{Mult} & \textbf{Power} & \textbf{Mult} & \textbf{Power} \\
\midrule
MFB & 10.874 & 8.397 & 0.329 & 0.271 \\
ECAPA-TDNN & 12.363 & 10.370 & 0.360 & 0.322 \\
GeMAPS & 11.883 & 9.467 & 0.354 & 0.297 \\
\midrule 
Average Probs & 8.251 & 7.764 & 0.284 & 0.252 \\
Average Logits & 7.698 & 7.660 & 0.266 & 0.240 \\
Dynamic Logits & 7.342 & 7.691 & 0.263 & 0.237 \\
Entropy & 7.533 & 7.758 & 0.264 & 0.242 \\
MetaLearner & 7.534 & 7.234 & 0.252 & 0.229 \\
\midrule 
DOVER-Lap & 10.610 & 7.940 & -- & -- \\
\bottomrule
\end{tabular}
\end{table}

For individual models, the results show a striking pattern: calibration in multilabel space consistently degrades performance compared to powerset calibration. Looking at Table~\ref{tab:calibration_probability_spaces}, individual models calibrated in Mult space show substantially worse DER than when calibrated in powerset space (MFB: 10.874\% vs 8.397\%, ECAPA-TDNN: 12.363\% vs 10.370\%, GeMAPS: 11.883\% vs 9.467\%). This degradation becomes even clearer when examining Table~\ref{tab:exp_1_combined}, which shows performance before and after calibration: for non-fine-tuned models, calibrating in powerset space improves DER (e.g., MFB: 10.381\% to 8.397\%), but the same models calibrated in multilabel space (Table~\ref{tab:calibration_probability_spaces}) actually perform worse than their uncalibrated versions (MFB: 10.874\% vs 10.381\% uncalibrated). This indicates that multilabel calibration not only fails to improve individual model performance but can actively harm it.

In contrast, fusion methods demonstrate more robustness to the calibration space choice. Interestingly, despite degrading individual model performance, multilabel calibration can still improve fusion results compared to using uncalibrated models. Comparing Table~\ref{tab:calibration_probability_spaces} with Table~\ref{tab:fusion_comparison}, fusion with multilabel-calibrated models (e.g., Average Probs: 8.251\% DER) outperforms fusion without calibration (8.425\% DER), suggesting that calibration normalizes prediction scales across models even in suboptimal spaces. 

However, the choice of calibration space affects DER and BCE differently for fusion methods. While BCE consistently improves with powerset calibration across all fusion methods without exception (e.g., Average Probs: 0.252 vs 0.284, Dynamic Logits: 0.237 vs 0.263, MetaLearner: 0.229 vs 0.252), the DER results show no clear pattern, with some methods favoring multilabel (Dynamic Logits: 7.342\% vs 7.691\%, Entropy: 7.533\% vs 7.758\%) and others favoring Power (Average Probs: 7.764\% vs 8.251\%, MetaLearner: 7.234\% vs 7.534\%). This discrepancy reinforces our earlier finding that optimizing calibration quality (BCE) does not necessarily align with optimizing diarization performance (DER).

\subsection{Effect of Calibration-Fusion Ordering}

This section explores whether calibration can be performed after combining base model predictions. Table~\ref{tab:cal_fusion_order} compares the performance of both ordering strategies for each fusion method: Calibrate-then-Fuse (C→F) and Fuse-then-Calibrate (F→C). Results are shown for non-fine-tuned and fine-tuned base models and compared with DOVER-Lap. Note that the F→C approach cannot be applied to DOVER-Lap, as it performs fusion at the hard decision level where frame-level speaker scores are no longer available. All configurations are fused in the multilabel space and calibrated in the powerset space.

\begin{table}[htbp]
\centering
\caption{DER (\%) comparison for calibration and fusion order on powerset space. Results shown for Calibrate-then-Fuse (C→F) and Fuse-then-Calibrate (F→C) strategies, with and without fine-tuning (FT).}
\label{tab:cal_fusion_order}
\resizebox{\columnwidth}{!}{%
\begin{tabular}{lccccc}
\toprule
 &  & \multicolumn{2}{c}{\textbf{DER (\%)}} & \multicolumn{2}{c}{\textbf{BCE}} \\
\cmidrule(r){3-4} \cmidrule(l){5-6}
\textbf{Method} & \textbf{Strategy}  & \textbf{No FT} & \textbf{FT} & \textbf{No FT} & \textbf{FT} \\
\midrule
\multirow{2}{*}{Average Probs} 
& C→F & 7.764 & 7.031 & 0.252 & 0.223\\
& F→C & 7.792 & 6.975 & 0.242 & 0.218 \\
\midrule
\multirow{2}{*}{Average Logits} 
& C→F & 7.660 & 7.030 & 0.240 & 0.210 \\
& F→C & 7.512 & 6.664 & 0.240 & 0.213 \\
\midrule
\multirow{2}{*}{Dynamic Logits} 
& C→F & 7.691 & 6.910 & 0.237 & 0.213 \\
& F→C & 7.458 & 6.543 & 0.239 & 0.217 \\
\midrule
\multirow{2}{*}{Entropy} 
& C→F & 7.758 & 7.030 & 0.242 & 0.217 \\
& F→C & 7.498 & 6.989 & 0.238 & 0.218 \\
\midrule
\multirow{2}{*}{MetaLearner} 
& C→F & 7.234 & 6.762 & 0.229 & 0.213 \\
& F→C & 7.202 & 6.651 & 0.233 & 0.214 \\
\midrule[\heavyrulewidth]
DOVER-Lap (Baseline)& C→F & 7.940 & 6.910 & -- & -- \\
\bottomrule
\end{tabular}
}
\end{table}

The results demonstrate that calibration after fusion (F→C) is not only feasible but often yields superior performance compared to the C→F approach. For fine-tuned models, all fusion methods except Average Probs show improvements with the F→C strategy. This F→C ordering also offers a significant computational advantage: only the single fused model requires calibration, rather than all $M$ base models in the C→F approach.

Comparing against the DOVER-Lap baseline, several fusion strategies already outperform it with the C→F approach. With the F→C strategy, performance improvements become more consistent---all methods except Average Probs surpass DOVER-Lap's 6.910\% DER for fine-tuned models. The best F→C result (6.543\% with Dynamic Logits) represents a substantial improvement over DOVER-Lap while requiring calibration of only a single combined model.

\subsection{DER Components and Calibration Quality Analysis}

\begin{figure*}[htbp!]
  \centering
  \includegraphics[width=0.9\linewidth]{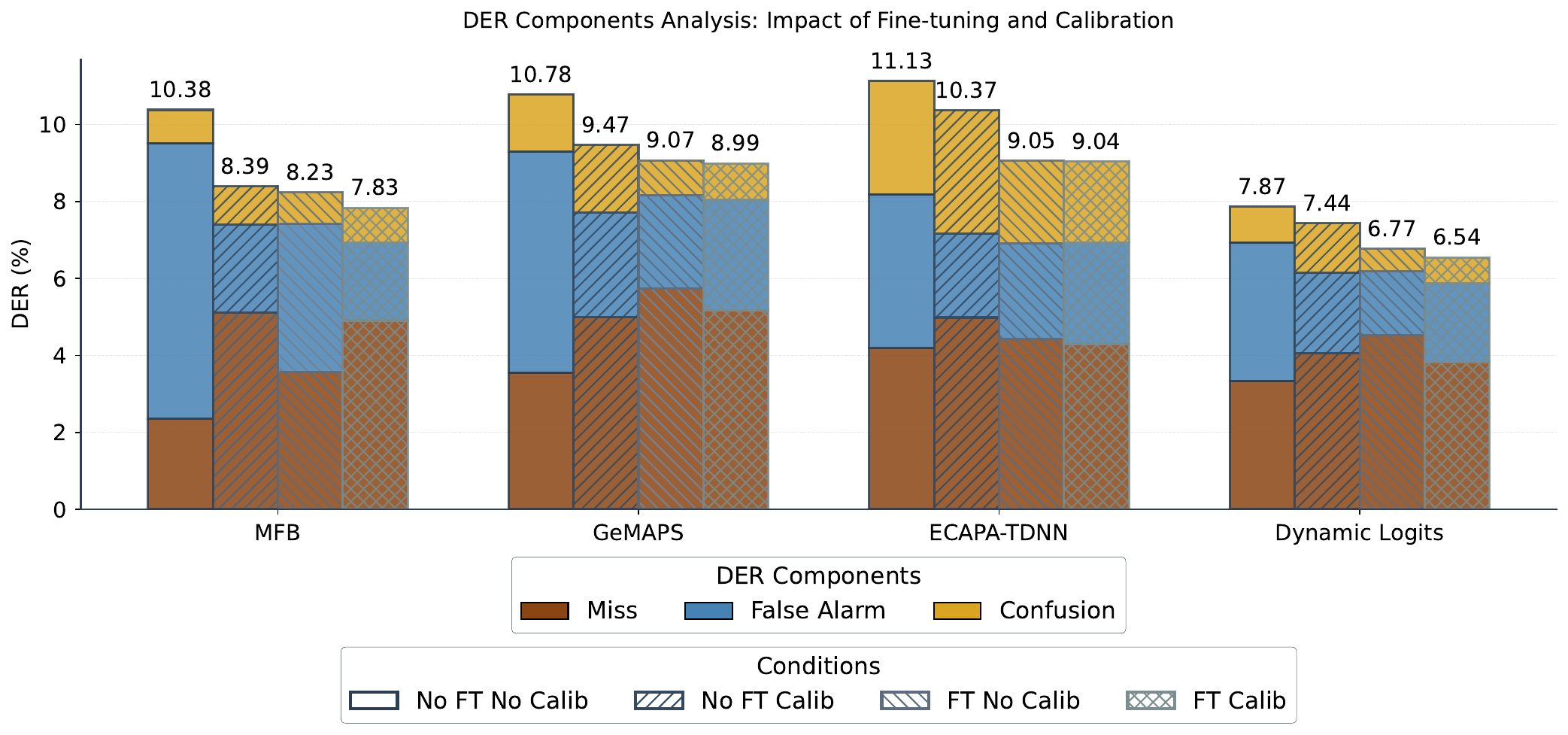}
  \caption{DER (\%) components for individual models and the best fusion method at different processing stages.}
  \label{fig:dercomponents}
\end{figure*}

\begin{figure}[h]
  \centering
  \includegraphics[width=\linewidth]{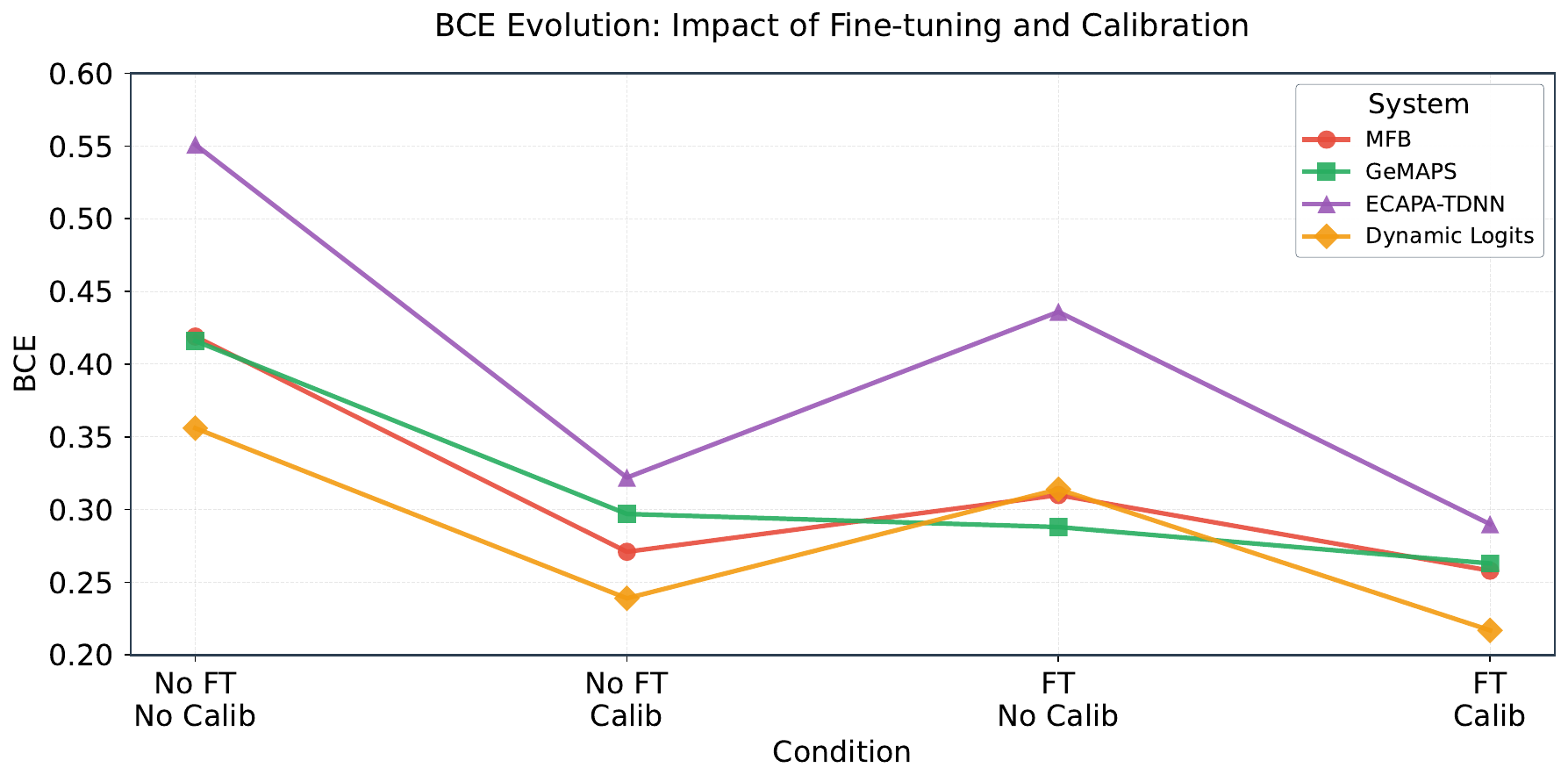}
  \caption{BCE for individual models and the best fusion method at different processing stages.}
  \label{fig:bce_evo}
\end{figure}

This section provides a visual analysis of how calibration, fine-tuning, and fusion affect DER error components (miss, false alarm, and confusion) and BCE at the system level. We compare individual base models against the best fusion method (Dynamic Logits) across four processing stages: no fine-tuning without calibration (No FT No Calib), no fine-tuning with calibration (No FT Calib), fine-tuning without calibration (FT No Calib), and fine-tuning with calibration (FT Calib). All results use joint calibration in the powerset space.

Figure~\ref{fig:dercomponents} reveals how calibration and fine-tuning affect error distribution across different DER components. The effect of calibration is consistent across all systems: calibration reduces false alarms while increasing miss errors, but the magnitude of false alarm reduction substantially exceeds the miss error increase, resulting in overall DER improvement. For example, MFB shows false alarms decreasing from 7.161\% to 2.275\% (a reduction of 4.886 percentage points) while miss errors increase from 2.351\% to 5.117\% (an increase of only 2.766 percentage points), yielding a net improvement in total DER. This asymmetric behavior suggests that calibration relaxes overconfident predictions, shifting decision boundaries toward more conservative thresholds that reduce spurious detections more effectively than they increase missed speech. Fine-tuning further enhances this effect by providing domain-adapted representations that work synergistically with calibration. The Dynamic Logits fusion consistently outperforms individual models across all conditions, demonstrating the benefits of combining calibrated predictions---particularly evident in the substantial reduction of confusion errors, where the fusion effectively resolves speaker identity ambiguities that individual models struggle with.

Figure~\ref{fig:bce_evo} shows BCE evolution across processing stages, revealing complementary effects of calibration and fusion. Calibration provides the largest BCE improvements for non-fine-tuned models. Notably, the Dynamic Logits fusion consistently achieves lower BCE than individual base models across nearly all processing stages, with the sole exception of the FT No Calib condition. This pattern demonstrates that combining predictions from diverse models improves prediction quality even without explicit calibration, as the fusion process naturally benefits from averaging complementary and partially uncorrelated model uncertainties. The combination of fine-tuning and calibration yields the best overall results, with Dynamic Logits achieving 0.217 BCE and 6.543\% DER, underscoring the synergistic benefits of domain adaptation, model diversity through fusion, and proper calibration.

The relationship between BCE and DER improvements reveals that better calibration typically leads to improved task performance. Across most systems and conditions, reducing BCE through calibration corresponds to DER reductions, demonstrating that well-calibrated confidence estimates support better diarization decisions. For instance, MFB achieves both substantial BCE improvement (0.419 to 0.271) and significant DER reduction (10.381\% to 8.397\%) through calibration alone. However, some interesting exceptions exist: ECAPA-TDNN shows minimal DER improvement (9.054\% to 9.036\%) despite substantial BCE reduction (0.436 to 0.290), suggesting that while its predictions become better calibrated, the improved confidence estimates do not translate to better speaker boundary decisions in this particular case. Similarly, comparing Dynamic Logits configurations shows that the non-fine-tuned calibrated version achieves better BCE (0.239) than the fine-tuned uncalibrated version (0.314), yet the latter yields superior DER (6.77\% vs 7.44\%), highlighting that domain adaptation provides complementary benefits beyond calibration. These observations indicate that while calibration quality and task performance are generally well-aligned, optimal diarization performance requires both well-calibrated predictions and domain-adapted representations. Dynamic Logits with both fine-tuning and calibration demonstrates this synergy, achieving the best overall results.

\section{Conclusions}
\label{sec:conclusion}

This work presents the first comprehensive framework for calibrating and fusing EEND systems at the probability level. Through experiments on CallHome, we demonstrate that proper calibration provides substantial improvements even for individual models (up to 19\% relative DER reduction), in some cases mitigating the absence of domain adaptation.

Our key findings proposes critical best practices for neural diarization: (1) Powerset representations with joint calibration consistently outperform independent per-speaker calibration (8.26\% vs 9.18\% DER), highlighting the importance of modeling speaker dependencies explicitly; (2) The Fuse-then-Calibrate strategy achieves superior performance (6.543\% DER, 5.2\% relative improvement over DOVER-Lap) while requiring calibration of only a single combined model; (3) Dynamic Logits fusion demonstrates the best performance across experimental conditions, effectively combining models at the logit level before softmax transformation.

We observe that calibration quality and task performance are not always aligned—improved BCE does not necessarily translate to improved DER. This finding highlights the need for task-specific calibration objectives in practical deployments. While we use a fixed threshold of 0.5 for speaker decisions in this work, the application of decision theory to optimize thresholds based on application-specific costs and well-calibrated probabilities remains an important direction for future work. Furthermore, calibration and fine-tuning prove complementary: calibration alone provides significant gains when domain data is unavailable, while their combination yields optimal results.

Future work should extend this framework to scenarios with more speakers, explore alternative proper scoring rules that may better correlate with DER optimization, and investigate applications in downstream tasks such as speaker-attributed automatic speech recognition. By demonstrating that probability-level techniques outperform hard-decision approaches while providing reliable confidence estimates, this work establishes the foundation for more effective neural diarization systems that fully exploit their probabilistic outputs.

\section*{Acknowledgments}

This research was supported by project PID2021-125943OB-I00 funded by MCIN/AEI/10.13039/501100011033 FEDER, UE and project PID2024-160789OB-I00 funded by MICIU/AEI/10.13039/501100011033 FEDER, UE and project SI4/PJI/2024-00237 (COSER-IA), Comunidad de Madrid.


 

\bibliographystyle{IEEEtran}
\bibliography{mybib}




\newpage

 



\vfill

\end{document}